\documentclass{article}
\usepackage{hello}

\usepackage[margin=1in]{geometry}

\usepackage{graphicx}
\usepackage{booktabs}
\usepackage{array}
\usepackage{amsmath}
\usepackage{bbm}
\usepackage{amssymb}
\usepackage{amsfonts}
\usepackage{multirow}
\usepackage{verbatim}
\usepackage{graphicx}
\usepackage{caption}
\usepackage{url}
\usepackage{longtable}
\usepackage{supertabular}
\usepackage{float}
\usepackage{enumitem}
\usepackage{tablefootnote}
\usepackage[round,semicolon]{natbib}
\usepackage{xcolor}
\usepackage{colortbl}
\usepackage{xspace}
\usepackage{textcomp}
\usepackage{makecell}
\usepackage{multirow}
\usepackage{lscape} 
\usepackage{siunitx}
\usepackage{tabularx}
\usepackage{comment}

\usepackage{amssymb}
\usepackage{pifont}
\usepackage{scrextend}
\usepackage{float}
\usepackage{array}
\usepackage{tgpagella}
\usepackage{latexsym}
\usepackage[T1]{fontenc}
\usepackage[utf8]{inputenc}
\usepackage{microtype}
\definecolor{mydarkblue}{rgb}{0,0.08,0.45}
\usepackage[colorlinks,citecolor=mydarkblue,urlcolor=mydarkblue,linkcolor=mydarkblue]{hyperref}
\usepackage{url}            
\usepackage{nicefrac}       
\usepackage{changepage}
\usepackage{xargs}          
\usepackage{wrapfig,lipsum,booktabs}
\usepackage{longtable}
\usepackage{caption}
\usepackage{subcaption}
\usepackage{endnotes}
\usepackage{tablefootnote} 

\usepackage{pgfplots}
\usetikzlibrary{pgfplots.groupplots}
\pgfplotsset{compat=1.3}
\usepackage{tikz}
\usetikzlibrary{patterns}
\usepackage[most]{tcolorbox}

\usepackage[capitalize,noabbrev]{cleveref}
\crefname{section}{Section}{\S\S}
\Crefname{section}{Section}{\S\S}
\crefname{table}{Table}{Tables}
\crefname{figure}{Figure}{Figures}
\crefname{algorithm}{Algorithm}{}
\crefname{equation}{eq.}{}
\crefname{appendix}{Appendix}{}
\crefformat{section}{Section #2#1#3}
\usepackage{multicol}
\usepackage{multirow}
\usepackage{bm}

\usepackage{tcolorbox}
\usepackage{titlesec}
\titleformat*{\section}{\large\bfseries}

\definecolor{battleshipgrey}{rgb}{0.3, 0.3, 0.3}
\definecolor{brilliantrose}{rgb}{1.0, 0.33, 0.64}
\definecolor{americanrose}{rgb}{1.0, 0.01, 0.24}
\definecolor{jweigreen}{rgb}{0,0.45,0.24}
\definecolor{bluegray}{rgb}{0.1, 0.1, 0.4}
\definecolor{ao(english)}{rgb}{0.0, 0.5, 0.0}
\definecolor{blanchedalmond}{rgb}{1.0, 0.92, 0.8}
\definecolor{atomictangerine}{rgb}{1.0, 0.6, 0.4}
\definecolor{chocolate(web)}{rgb}{0.82, 0.41, 0.12}
\definecolor{bananayellow}{rgb}{1.0, 0.88, 0.21}
\definecolor{goldenbrown}{rgb}{0.6, 0.4, 0.08}
\definecolor{aliceblue}{rgb}{0.94, 0.97, 1.0}
\definecolor{beige}{rgb}{0.96, 0.96, 0.86}
\definecolor{babyblue}{rgb}{0.54, 0.81, 0.94}
\definecolor{camel}{rgb}{0.76, 0.6, 0.42}
\definecolor{cinnamon}{rgb}{0.82, 0.41, 0.12}
\definecolor{deepskyblue}{rgb}{0.0, 0.75, 1.0}
\definecolor{frenchblue}{rgb}{0.0, 0.45, 0.73}
\definecolor{classicrose}{rgb}{0.98, 0.8, 0.91}
\definecolor{frenchrose}{rgb}{0.96, 0.29, 0.54}
\definecolor{frenchlilac}{rgb}{0.53, 0.38, 0.56}
\definecolor{frenchbeige}{rgb}{0.65, 0.48, 0.36}
\definecolor{verylightgreen}{RGB}{240, 255, 235}
\definecolor{verylightred}{RGB}{255, 235, 235}
\definecolor{verylightyellow}{RGB}{255, 254, 235}
\definecolor{dt}{gray}{0.7}

\definecolor{forestgreen}{HTML}{2e7d43}
\definecolor{color1}{HTML}{FF9999}
\definecolor{color2}{HTML}{FF6666}
\definecolor{color3}{HTML}{FF3333}
\definecolor{color4}{HTML}{E60000}
\definecolor{color5}{HTML}{B30000}
\definecolor{color6}{HTML}{8CD98C}
\definecolor{color7}{HTML}{53c653}
\definecolor{color8}{HTML}{39ac39}
\definecolor{color9}{HTML}{2d862d}
\definecolor{color10}{HTML}{206020}
\definecolor{color11}{HTML}{cca300}

\usepackage{minitoc}

\title{
\textbf{
Qwen2-Audio Technical Report}
}

\author{
\large{}
Yunfei Chu$^*$$^{\dag}$ \hspace{6mm} Jin Xu$^*$$^{\dag}$ \hspace{6mm} Qian Yang$^*$ \hspace{6mm} Haojie Wei \hspace{6mm}\\
Xipin Wei \hspace{6mm} Zhifang Guo \hspace{6mm} Yichong Leng \hspace{6mm} Yuanjun Lv \hspace{6mm} Jinzheng He \hspace{6mm} \\
Junyang Lin \hspace{6mm} Chang Zhou$^{\dag}$ \hspace{6mm} Jingren Zhou
\\
\large{}
Qwen Team, Alibaba Group
\\
\\
\small{}
Code \& Demo \& Models: \ \ \url{https://github.com/QwenLM/Qwen2-Audio}
}

\date{}

\begin{document}

\doparttoc 
\faketableofcontents 

\maketitle

\begin{abstract}
\noindent

We introduce the latest progress of Qwen-Audio, a large-scale audio-language model called Qwen2-Audio, which is capable of accepting various audio signal inputs and performing audio analysis or direct textual responses with regard to speech instructions. In contrast to complex hierarchical tags, we have simplified the pre-training process by utilizing natural language prompts for different data and tasks, and have further expanded the data volume. 
We have boosted the instruction-following capability of Qwen2-Audio and implemented two distinct audio interaction modes for voice chat and audio analysis. 
In the voice chat mode, users can freely engage in voice interactions with Qwen2-Audio without text input. In the audio analysis mode, users could provide audio and text instructions for analysis during the interaction. Note that we do not use any system prompts to switch between voice chat and audio analysis modes. Qwen2-Audio is capable of intelligently comprehending the content within audio and following voice commands to respond appropriately. For instance, in an audio segment that simultaneously contains sounds, multi-speaker conversations, and a voice command, Qwen2-Audio can directly understand the command and provide an interpretation and response to the audio.
Additionally, DPO has optimized the model's performance in terms of factuality and adherence to desired behavior. According to the evaluation results from AIR-Bench, Qwen2-Audio outperformed previous SOTAs, such as Gemini-1.5-pro, in tests focused on audio-centric instruction-following capabilities. Qwen2-Audio is open-sourced with the aim of fostering the advancement of the multi-modal language community.
\end{abstract}

{\let\thefootnote\relax\footnotetext{$^*$Equal contribution, $^\dag$Corresponding author}}

\begin{figure*}[t!]
\centering
\includegraphics[width=12cm]{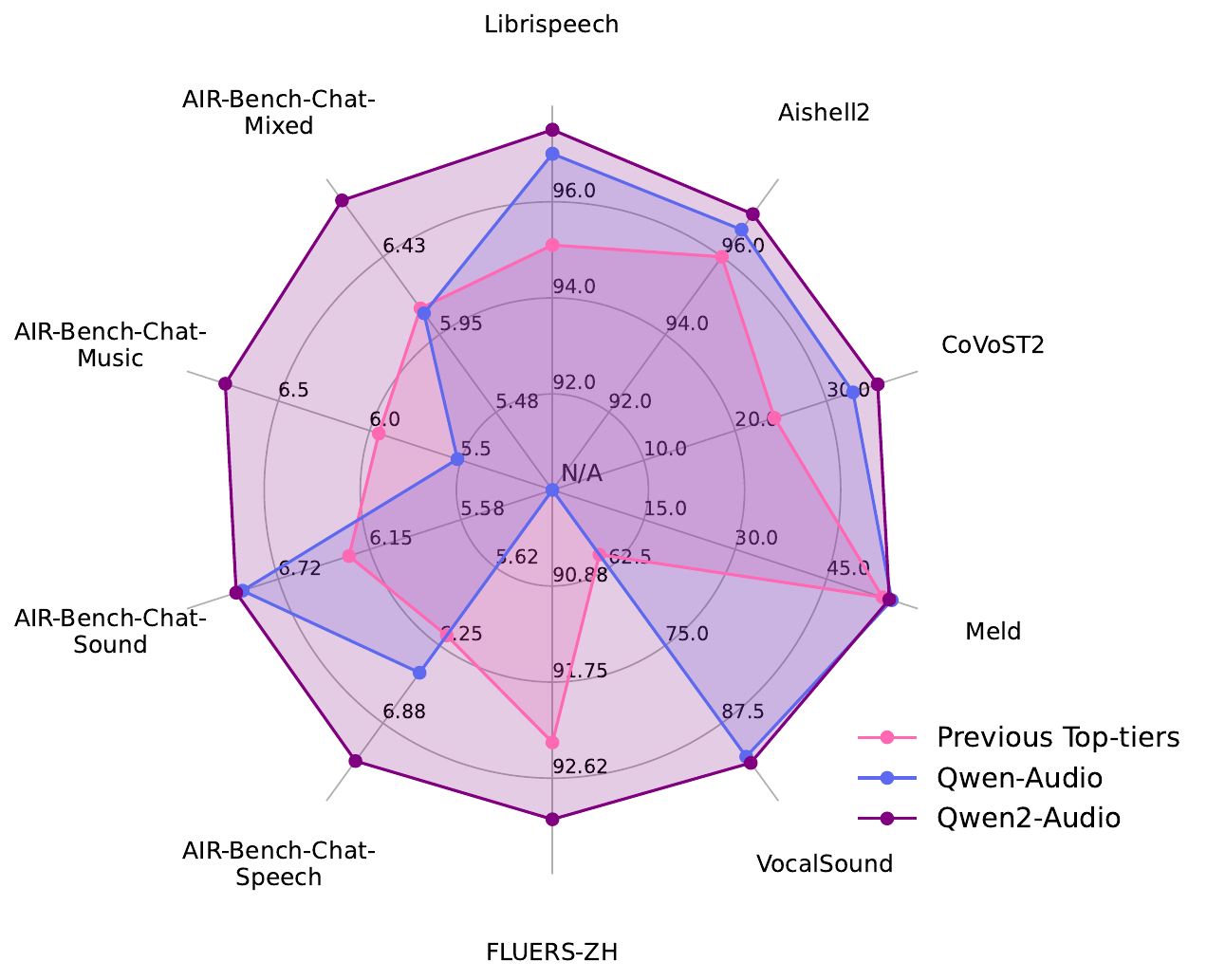}
\caption{Performance of Qwen2-Audio, Qwen-Audio and previous top-tiers from LALMs such as SpeechT5~\citep{ao2021speecht5}, SpeechNet~\citep{chen2021speechnet}, SpeechLLaMA~\citep{speechllama}, SALMONN~\citep{anonymous2023salmonn}, Whisper~\citep{Whisper} Pengi~\citep{Pengi}, and SpeechVerse~\citep{das2024speechverse}. We demonstrate the test set results across the 10 datasets covering Automatic Speech Recognition~(ASR), Speech-to-Text Translation~(S2TT), Speech Emotion Recognition~(SER), Vocal Sound Classification~(VSC), and instruction-following benchmark ~\citep{Yang2024AIRBenchBL}. The results of ASR datasets, such as Librispeech and Aishell2 refer to 1 - WER\%. The results of CoVoST2 is the average BLEU score of seven translation directions (en-de, de-en, en-zh, zh-en, es-en, fr-en and it-en). The results of the AIR-Bench chat benchmark encompass four dimensions: speech, sound, music, and mixed. Scores for each dimension are automatically assessed by GPT-4, with values ranging from 0 to 10. Qwen2-Audio achieves remarkable performance without requiring any task-specific fine-tuning, surpassing its counterparts.}
\label{img:radar}
\end{figure*}

\section{Introduction}
Audio serves as a crucial medium for interaction and communication among humans and other living beings, carrying rich information content. A comprehensive understanding of various forms of audio signals is paramount to achieving Artificial General Intelligence (AGI). Recently, significant advancements have been made in the development of large audio-language models (LALMs)~\citep{chu2023qwen, das2024speechverse, kong2024audio, anonymous2023salmonn, openai2024gpt4-o}, demonstrating remarkable achievements in comprehending diverse speech signals, performing speech signal analysis, and complex reasoning.

In this report, we develop Qwen2-Audio, with a primary focus on enhancing its instruction-following capabilities. Qwen2-Audio is a Large Audio-Language Model (LALM) designed to process both audio and text inputs to generate textual outputs. Compared to previous models, Qwen2-Audio significantly scales up the training dataset. To reduce the gap between pre-training and post-training stages, we simplify the pre-training process by directly using natural language prompts for various data and tasks, as illustrated in figure~\ref{img:framework}. Following the practices in Large Language Models (LLMs)~\citep{gpt4, qwen7b}, we further conduct instruction tuning and direct preference optimization to align the model's outputs with human preferences.

Qwen2-Audio operates in two distinct modes: Audio Analysis and Voice Chat. These two modes are differentiated by their functionality, but there is no need for users to distinguish between them during use.
In the audio analysis mode, users can leverage Qwen2-Audio to analyze a diverse range of audio types, including speech, sound, music, or various mixed audio forms. Commands can be issued either through audio or text, and Qwen2-Audio will autonomously discern the command segments within the audio. Conversely, in voice chat mode, users can interact with Qwen2-Audio as if it were a conversational agent, engaging in unrestricted dialogue. Audio interaction is available, and users can switch to text interaction at any moment they choose.
For instance, if a user inputs an audio clip where the initial part is the sound of typing on a keyboard, followed by the user asking "What is this sound?" in spoken language, Qwen2-Audio is expected to respond directly with "This is the sound of a keyboard."

As shown in Figure~\ref{img:radar}, extensive evaluation demonstrates that Qwen2-Audio, without any task-specific fine-tuning, outperforms previous LALMs across a diverse range of tasks. Among them, Qwen2-Audio achieves state-of-the-art performance on the test set of Aishell2, FLUERS-zh, VocalSound and AIR-Bench chat benchmark.

\section{Methodology}
\label{method}

\begin{figure*}[t!]
\centering
\includegraphics[width=15cm]{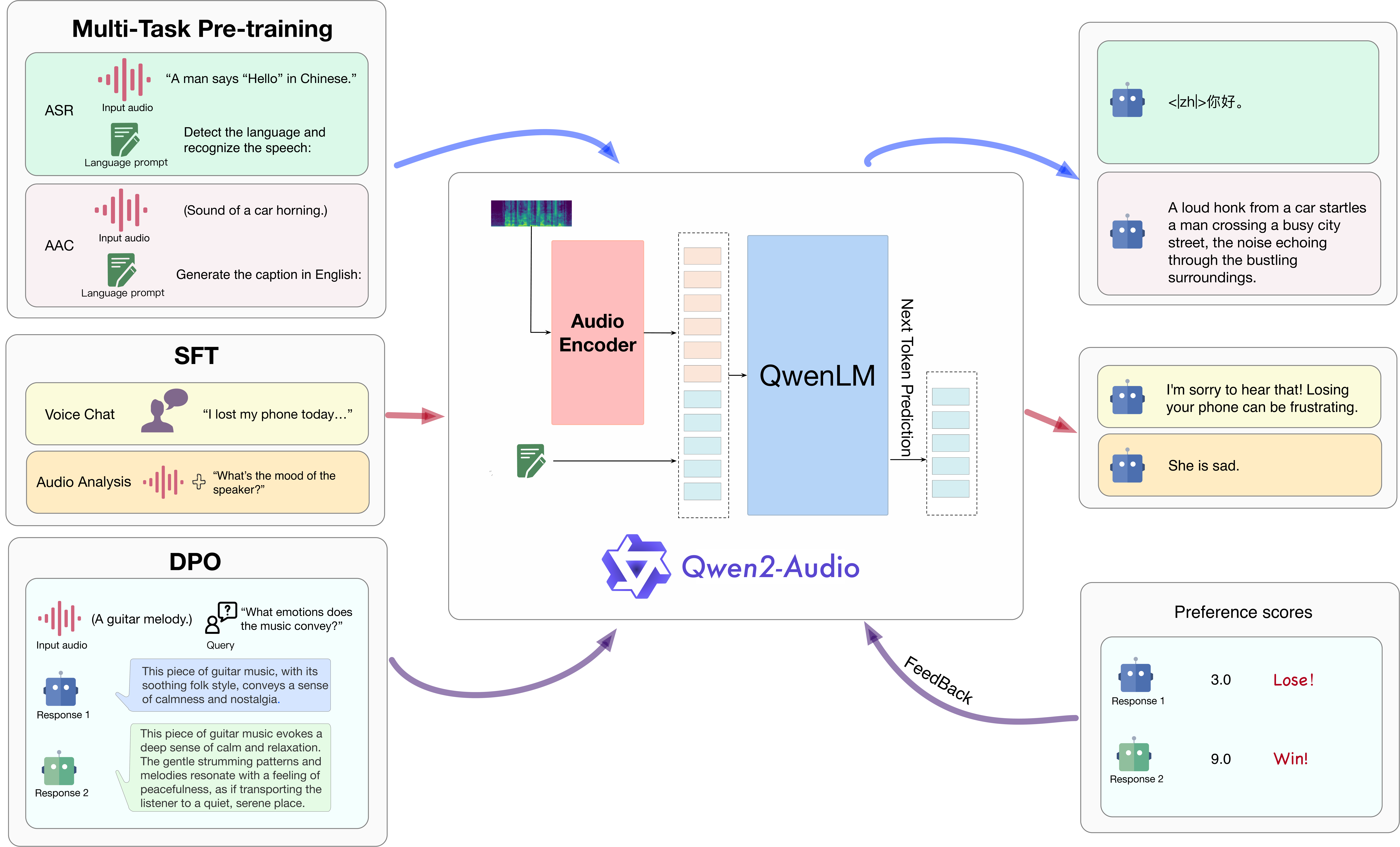}
   \caption{The overview of three-stage training process of Qwen2-Audio.}
\label{img:framework}
\end{figure*}

\paragraph{Model Architecture}\label{exp:model_arch}
The training process of Qwen2-Audio is depicted in Figure~\ref{img:framework}, which contains an audio encoder and a large language model. Given the paired data $(\bm{a}, \bm{x})$, where the $\bm{a}$ and $\bm{x}$ denote the audio sequences and text sequences, the training objective is to maximize the next text token probability as
\begin{equation}
    \mathcal{P}_{\theta}(x_t|\bm{x}_{<t}, \text{Encoder}_{\phi}(\bm{a})),
\end{equation}
conditioning on audio representations and previous text sequences $\bm{x}_{<t}$, where $\theta$ and $\phi$ denote the trainable parameters of the LLM and audio encoder respectively. 

Different from Qwen-Audio, the initialization of the audio encoder of Qwen2-Audio is based on the Whisper-large-v3 model~\citep{Whisper}. 
To preprocess the audio data, we resamples it to a frequency of 16kHz and converts the raw waveform into 128-channel mel-spectrogram using a window size of 25ms and a hop size of 10ms. Additionally, a pooling layer with a stride of two is incorporated to reduce the length of the audio representation. As a result, each frame of the encoder output approximately corresponds to a 40ms segment of the original audio signal. Qwen2-Audio still incorporates the large language model Qwen-7B~\citep{bai2023qwen} as its foundational component. The total parameters of Qwen2-Audio is 8.2B parameters.

\paragraph{Pre-training}\label{exp:multitask_train}
At the pre-training stage, we replace the hierarchical tags~\citep{chu2023qwen} with the natural language prompts. As shown in Figure~\ref{img:framework}. We find that using language prompts can improve better generalization ability and better instruction following ability.

\begin{figure}[ht]
\centering
\includegraphics[width=0.5\linewidth]{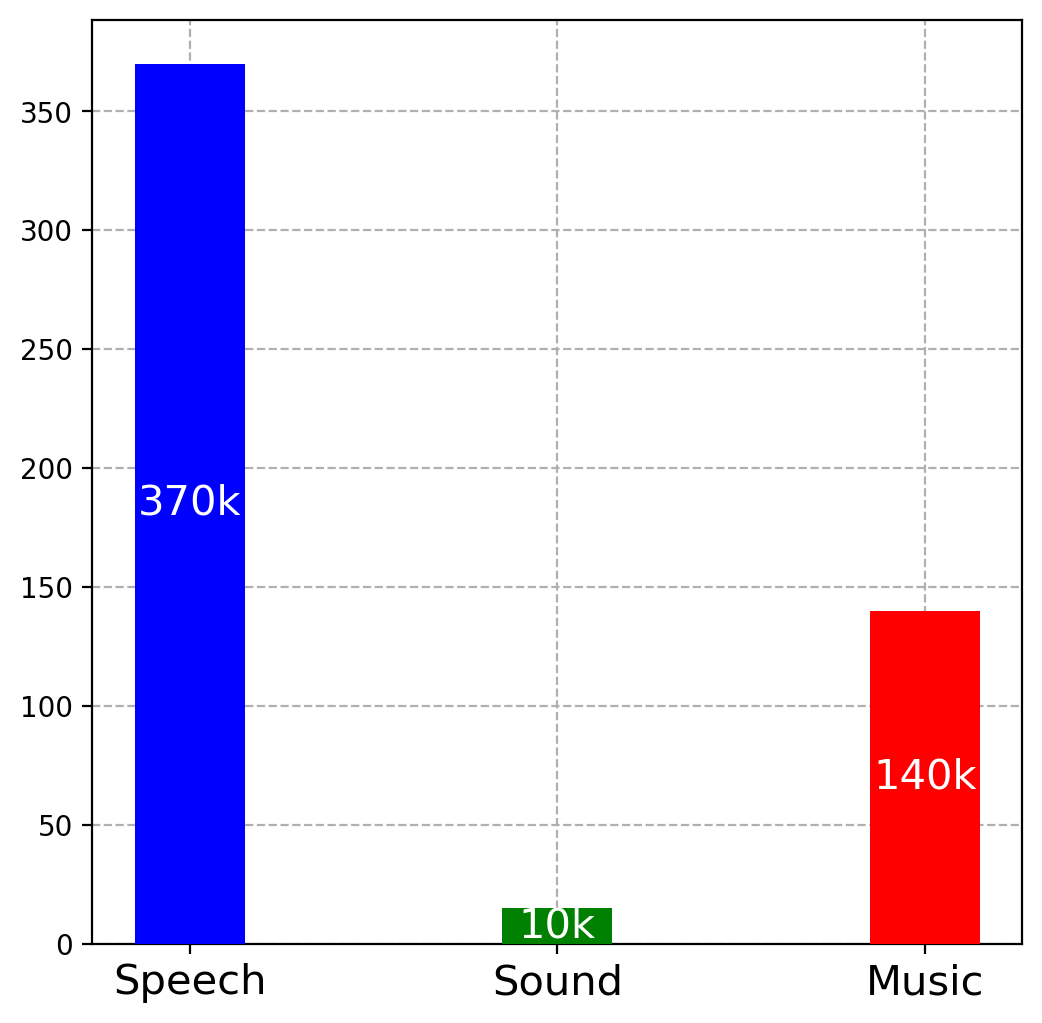}
\caption{Statistics (hours) of pre-training dataset.}
\label{Statistics of pre-training dataset}
\end{figure}

\paragraph{Supervised Fine-tuning}\label{exp:sft}
The thorough pretraining of Qwen2-Audio has equipped the model with a comprehensive understanding of audio content. Building upon this, we employ instruction-based fine-tuning techniques to improve the ability of the model to align with human intent, resulting in an interactive chat model.
Our prelimilary study emphasizes the critical influence of the quality and complexity of SFT data on the model's performance. Accordingly, a meticulously curated set of high-quality SFT data was collected, with rigorous quality control procedures implemented.

We consider two distinct modes for human interactions:

\begin{itemize}
    \item \textbf{Audio Analysis}: 
    In the audio analysis mode, users are afforded the flexibility to have Qwen2-Audio analyze a diverse array of audio. User instructions can be given either through audio or text.
    
    This mode is often used for offline analysis of audio files.
    \item \textbf{Voice Chat}: In the voice chat mode, users are encouraged to engage in voice conversations with Qwen2-Audio, asking a wide range of questions. Please feel free to consider it your voice chat assistant.
    
    This mode is often used for online interaction with LALMs.
\end{itemize}
For consistency and model uniformity, both interaction modes were jointly trained, thus users will not experience mode differentiation during use, nor is it necessary to switch between different modes using separate system prompts. The two modes are seamlessly integrated in actual use.

\paragraph{Direct Preference Optimization}\label{exp:dpo}
We employ DPO~\citep{rafailov2024direct} to further optimize models to follow human preferences. By obtaining the dataset $\mathcal{D}$ with the triplet data $(\bm{x}, \bm{y_w}, \bm{y_l})$, where $\bm{x}$ is the input sequence with input audio, and $\bm{y_w}$ and $\bm{y_l}$ are the human-annotated good and bad responses respectively, we optimize the model $\mathcal{P}_\theta$ as follows:
\begin{equation}
    \mathcal{L}_{\text{DPO}}(\mathcal{P}_\theta; \mathcal{P}_{\text{ref}}) = -\mathbb{E}_{(\bm{x}, \bm{y_w}, \bm{y_l}) \sim \mathcal{D}} \left[ \log \sigma \left( \beta \log \frac{\mathcal{P}_\theta(\bm{y_w} \mid \bm{x})}{\mathcal{P}_{\text{ref}}(\bm{y_w} \mid \bm{x})} - \beta \log \frac{\mathcal{P}_\theta(\bm{y_l} \mid \bm{x})}{\mathcal{P}_{\text{ref}}(\bm{y_l} \mid \bm{x})} \right) \right],
\end{equation}
where $\mathcal{P}_{\text{ref}}$ denotes the reference model initialized with $\mathcal{P}_\theta$, $\sigma$ represents sigmoid function and $\beta$ is a hyperparameter. 
Figure~\ref{img:framework} illustrates the three-stage training process of Qwen2-Audio.

\begin{table}[t!]
\centering
\caption{Summary of Evaluation Benchmarks for Qwen2-Audio.}
\vspace{-2mm}
\resizebox{\textwidth}{!}{%
\begin{tabular}{lcccc}
\toprule
\textbf{Task}                  & \textbf{Description}                   & \textbf{Dataset}                    & \textbf{Split} & \textbf{Metric} \\
\midrule
\multirow{4}{*}{ASR} & \multirow{4}{*}{Automatic Speech Recognition} & Fleurs~\citep{Conneau2022FLEURSFL}  & dev | test     & \multirow{4}{*}{WER}    \\
                     &                                              & Aishell2~\citep{aishell2}            & test           &                      \\
                     &                                              & Librispeech~\citep{Librispeech}      & dev | test     &                      \\
                     &                                              & Common Voice~\citep{commonvoice:2020}& dev | test     &                      \\
\midrule
S2TT                 & Speech-to-Text Translation                   & CoVoST2~\citep{CoVoST2}              & test           & BLEU \footnotemark~\citep{papineni2002bleu} \\
\midrule
SER                  & Speech Emotion Recognition                   & Meld~\citep{Meld}                    & test           & ACC                 \\
\midrule
VSC                  & Vocal Sound Classification                   & VocalSound~\citep{VocalSound}        & test           & ACC                 \\
\midrule
\multirow{4}{*}{\begin{tabular}[l]{@{}l@{}}AIR-Bench\\\citep{Yang2024AIRBenchBL}\end{tabular}}  & Chat-Benchmark-Speech & \begin{tabular}[c]{@{}c@{}}Fisher~\citep{cieri2004fisher}\\ SpokenWOZ~\citep{si2023spokenwoz}\\ IEMOCAP~\citep{si2023spokenwoz}\\ Common voice~\citep{commonvoice:2020}\end{tabular} & dev | test & GPT-4 Eval \\ \cmidrule{2-5} 
 & Chat-Benchmark-Sound & Clotho~\citep{Clotho} & dev | test & GPT-4 Eval \\ \cmidrule{2-5} 
 & Chat-Benchmark-Music & MusicCaps~\citep{agostinelli2023musiclm} & dev | test & GPT-4 Eval \\ \cmidrule{2-5} 
 & Chat-Benchmark-Mixed-Audio & \begin{tabular}[c]{@{}c@{}}Common voice~\citep{commonvoice:2020}\\ AudioCaps~\citep{kim2019audiocaps}\\ MusicCaps~\citep{agostinelli2023musiclm}\end{tabular} & dev | test & GPT-4 Eval \\ 
\bottomrule         
\end{tabular}
}
\label{tab:evaluation}
\end{table}
\footnotetext{https://github.com/mjpost/sacrebleu}

\section{Experiments}
\label{sec:experiments}

\subsection{Evaluation}
In practice, we have found that many previous test datasets are highly limited and cannot adequately reflect performance in real-world scenarios, such as some SLU (Spoken Language Understanding) and SER (Speech Emotion Recognition) datasets. Therefore, we mainly evaluated performance directly on AIR-Bench. We discovered that the scores from AIR-Bench align more closely with the actual user interaction experience. Meanwhile,
in order to assess the universal understanding capabilities of Qwen2-Audio, as shown in Table~\ref{tab:evaluation}, we still perform a comprehensive evaluation that encompasses various tasks, namely Automatic Speech Recognition (ASR), Speech-to-Text Translation (S2TT), Speech Emotion Recognition (SER), Vocal Sound Classification (VSC). The evaluation is conducted across 13 datasets. The evaluation datasets are rigorously excluded from the training data to avoid data leakage. The  models we compare include open-source models and callable APIs, such as Gemini.

\begin{table}[t!]
\centering
\caption{The results of Automatic Speech Recognition~(ASR), Speech-to-Text Translation~(S2TT), Speech Emotion Recognition~(SER), Vocal Sound Classification~(VSC), and AIR-Bench chat benchmark. Note that for Qwen2-Audio, the results for Fleurs are zero-shot, whereas the results for Common Voice are not zero-shot.}
\vspace{-2mm}

\resizebox{\textwidth}{!}{%
\begin{tabular}{lcccc}
\toprule
\multirow{2}{*}{\textbf{Task}} & \multirow{2}{*}{\textbf{Dataset} } & \multirow{2}{*}{\textbf{Model}} & \multicolumn{2}{c}{\textbf{Performance} }  \\ \cmidrule(l){4-5} &  &  & \textbf{Metrics}  & \textbf{Results} \\ \midrule \multirow{16}{*}{ASR} & \multirow{7}{*}{\begin{tabular}[c]{@{}c@{}}\textbf{Librispeech}\\ \textit{dev-clean} | \textit{dev-other} | \\ \textit{test-clean} | \textit{test-other} \end{tabular}}     & SpeechT5~\citep{ao2021speecht5}    & \multirow{6}{*}{WER~$\downarrow$}    & 2.1 | 5.5 | 2.4 | 5.8   \\ &  & SpeechNet~\citep{chen2021speechnet}  & & - | - | 30.7 | -   \\ &   & SLM-FT~\citep{SLM} & & - | - | 2.6 | 5.0  \\ &  & SALMONN~\citep{anonymous2023salmonn}   & & - | - | 2.1 | 4.9  \\ & & SpeechVerse~\citep{das2024speechverse} & & - | - | 2.1 | 4.4   \\ & & Qwen-Audio~\citep{chu2023qwen} & & 1.8 | 4.0 | 2.0 | 4.2   \\ & & Qwen2-Audio & & \textbf{1.3} | \textbf{3.4} | \textbf{1.6} | \textbf{3.6}   \\
\cmidrule(l){2-5}  & \multirow{2}{*}{\begin{tabular}[c]{@{}c@{}}\textbf{Common Voice 15} \\ \textit{en} | \textit{zh} | \textit{yue} | \textit{fr} \end{tabular}}  & Whisper-large-v3~\citep{Whisper}  & \multirow{2}{*}{WER~$\downarrow$} & 9.3 | 12.8 | 10.9 | 10.8   \\ & & Qwen2-Audio  & & \textbf{8.6} | \textbf{6.9} | \textbf{5.9} | \textbf{9.6} \\ \cmidrule(l){2-5} 
 & \multirow{2}{*}{\begin{tabular}[c]{@{}c@{}}\textbf{Fleurs} \\ \textit{zh}  \end{tabular}}        & Whisper-large-v3~\citep{Whisper}   &  \multirow{2}{*}{WER~$\downarrow$}   & 7.7  \\
  &  & Qwen2-Audio & & \textbf{7.5}  \\          
\cmidrule(l){2-5} 
   & \multirow{4}{*}{\begin{tabular}[c]{@{}c@{}}\textbf{Aishell2} \\ \textit{Mic} | \textit{iOS} | \textit{Android} \end{tabular}}        & MMSpeech-base~\citep{mmspeech}              &     \multirow{4}{*}{WER~$\downarrow$}              & 4.5 | 3.9 | 4.0                \\
                                                  &                           & Paraformer-large~\citep{FunASR}          &                      & - | \textbf{2.9} | -                 \\
                                                  
                                                  &                           & Qwen-Audio~\citep{chu2023qwen}           &                                         & 3.3 | 3.1 | 3.3                 \\
                                                  &                           & Qwen2-Audio           &                                         & \textbf{3.0} | 3.0 | \textbf{2.9}                 \\  
                                                  \midrule
\multirow{9}{*}{S2TT}    & \multirow{5}{*}{\begin{tabular}[c]{@{}c@{}}\textbf{CoVoST2} \\ \textit{en-de} | \textit{de-en} | \\ \textit{en-zh} | \textit{zh-en} \end{tabular}}     & SALMONN~\citep{anonymous2023salmonn}    & \multirow{5}{*}{BLEU~$\uparrow$}                    & 18.6 | - | 33.1 | -                \\
                                                  &                              & SpeechLLaMA~\citep{speechllama}                &                                         & - | 27.1 | - | 12.3            \\
                                                  
                                                  &                         & BLSP~\citep{wang2023blsp}            &                                         & 14.1 | - | - | -            \\
                                                  &                               & Qwen-Audio~\citep{chu2023qwen}                &                                         & 25.1 | 33.9 | 41.5 | 15.7            \\
                                                  &                               & Qwen2-Audio                &                                         & \textbf{29.9} | \textbf{35.2} | \textbf{45.2} | \textbf{24.4}       \\
\cmidrule(l){2-5}
                                                  & \multirow{3}{*}{\begin{tabular}[c]{@{}c@{}}\textbf{CoVoST2} \\ \textit{es-en} | \textit{fr-en} |  \textit{it-en} |  \end{tabular}}
                                                  & SpeechLLaMA~\citep{speechllama}              &    \multirow{3}{*}{BLEU~$\uparrow$}              & 27.9 | 25.2 | 25.9                \\
                                                  & & Qwen-Audio~\citep{chu2023qwen}                   &                                         & {39.7} | \textbf{38.5} | {36.0}                 \\ 
                                                  &                               & Qwen2-Audio                &                                         & \textbf{40.0} | \textbf{38.5} | \textbf{36.3}    \\

                                                  \midrule

\multirow{3}{*}{SER} & \multirow{3}{*}{\textbf{Meld}}           & WavLM-large~\citep{wavlm}            & \multirow{3}{*}{ACC~$\uparrow$}                    & 0.542                 \\
                                                  &                                 & Qwen-Audio~\citep{chu2023qwen}                    &                                         & \textbf{0.557}                 \\ 
                                                  &                                 & Qwen2-Audio                    &                                         & {0.553}                 \\ \midrule
\multirow{4}{*}{VSC}       & \multirow{4}{*}{\textbf{VocalSound}}     & CLAP~\citep{CLAP}                   & \multirow{4}{*}{ACC~$\uparrow$}                    & 0.4945                \\
                                                  &                                 & Pengi~\citep{Pengi}                  &                                         & 0.6035                \\
                                                  &                                 & Qwen-Audio~\citep{chu2023qwen}                    &                                         & 0.9289                 \\
                                                  &                                 & Qwen2-Audio                   &                                         & \textbf{0.9392}                \\ \midrule                                            
\begin{tabular}[l]{@{}l@{}}AIR-Bench \\\citep{Yang2024AIRBenchBL}\end{tabular} & \begin{tabular}[c]{@{}c@{}}\textbf{Chat Benchmark}\\\textit{Speech} | \textit{Sound} | \\  \textit{Music} | \textit{Mixed-Audio}\end{tabular} & \begin{tabular}[c]{@{}c@{}}SALMONN~\citep{anonymous2023salmonn}\\ BLSP~\citep{wang2023blsp}\\ Pandagpt~\citep{su2023pandagpt}\\ Macaw-LLM~\citep{Macaw-LLM}\\ SpeechGPT~\citep{speechgpt}\\ Next-gpt~\citep{NextGPT}\\ Qwen-Audio~\citep{chu2023qwen}
\\
Gemini-1.5-pro~\citep{reid2024gemini}

\\ Qwen2-Audio\end{tabular} & GPT-4~$\uparrow$ & \begin{tabular}[c]{@{}c@{}}6.16 | 6.28 | 5.95 | 6.08\\ 6.17 | 5.55 | 5.08 | 5.33\\ 3.58 | 5.46 
| 5.06 | 4.25\\ 0.97 | 1.01 | 0.91 | 1.01\\ 1.57 | 0.95 | 0.95 | 4.13\\ 3.86 | 4.76 | 4.18 | 4.13\\ 
6.47 | 6.95 | 5.52 | 6.08 \\
6.97 | 5.49 | 5.06 | 5.27
\\ \textbf{7.18} | \textbf{6.99} | \textbf{6.79} | \textbf{6.77}\end{tabular} \\ 
\bottomrule

\end{tabular}%
}

\label{tab:audio_analysis_table}
\end{table}

\subsection{Main Results}
In this section, we present a comprehensive evaluation of the Qwen2-Audio model, assessing its performance across various tasks without any task-specific fine-tuning. We begin by examining its English Automatic Speech Recognition (ASR) results, as depicted in Table~\ref{tab:audio_analysis_table}, where Qwen2-Audio exhibits superior performance compared to previous multi-task learning models. Specifically, it achieves a 1.6\% and 3.6\% WER on the librispeech test-clean and test-other datasets, respectively. Compared with Whisper-large-v3 on Fleurs zh subset, we achieve better results than Whisper-large-v3. One point to note is that Qwen2-Audio is not evaluated in a zero-shot manner on the Common Voice 15 dataset, whereas Whisper’s results are obtained in a zero-shot fashion. However, on the Fleurs dataset, both Qwen2-Audio and Whisper are evaluated in a zero-shot manner.
Furthermore, we evaluate Qwen2-Audio's speech translation performance on the CoVoST2 dataset. The results reveal that Qwen2-Audio outperforms the baselines by a substantial margin across all seven translation directions. For sound, we analyze the performance of Qwen2-Audio on SER, and VSC, as summarized in Table~\ref{tab:audio_analysis_table}. Across these tasks, Qwen2-Audio consistently outperforms the baselines by a significant margin. 

Lastly, to objectively evaluate the chat capabilities of Qwen2-Audio, we measured its performance on the chat benchmark of the AIR-Bench~\citep{Yang2024AIRBenchBL}. Note that since Gemini-1.5~\citep{reid2024gemini}\footnote{\url{https://console.cloud.google.com/vertex-ai/generative/multimodal/create}} cannot correctly return some test samples due to its SAFETY reasons during testing, the number of samples of Gemini-1.5 on AIR-Bench-chat has been reduced by about 1/5. As shown in table~\ref{tab:audio_analysis_table}, Qwen2-Audio demonstrates state-of-the-art (SOTA) instruction-following capabilities across speech, sound music and mixed-Audio subsets. It shows substantial improvements compared to Qwen-Audio and significantly outperforms other LALMs.

\section{Cases}
\label{case_study}
Here, we present a part of cases to illustrate Qwen2-Audio's audio-based interaction capabilities. For more impressive examples, please refer to \url{https://github.com/QwenLM/Qwen2-Audio}.

\begin{figure*}[htbp]
\centering
\includegraphics[ width=0.7\textwidth]
    {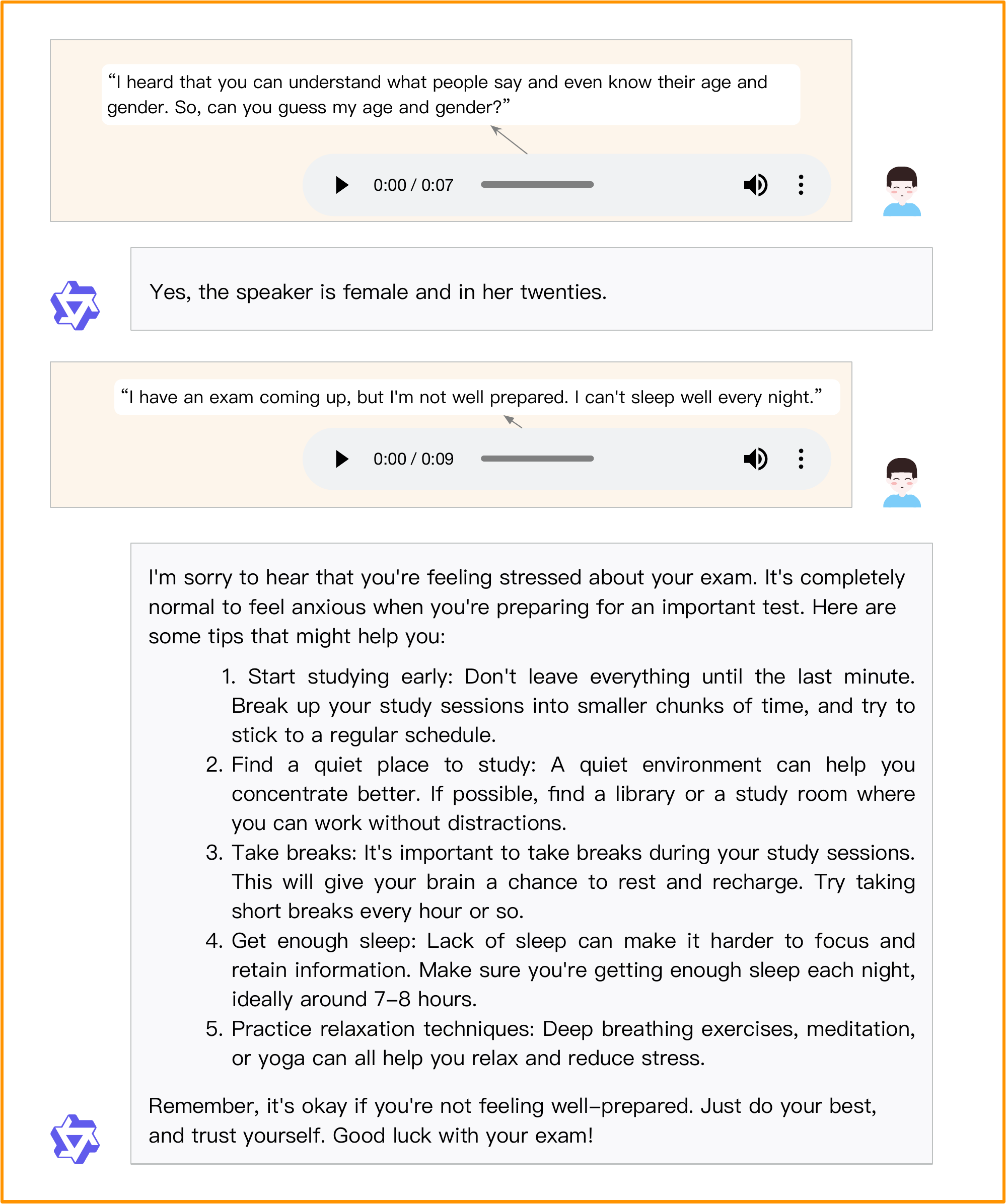}
\caption{Example showing Qwen2-Audio’s capability in free chat around speech.}
\end{figure*}

\begin{figure*}[htbp]
\centering
\includegraphics[width=0.7\textwidth]
    {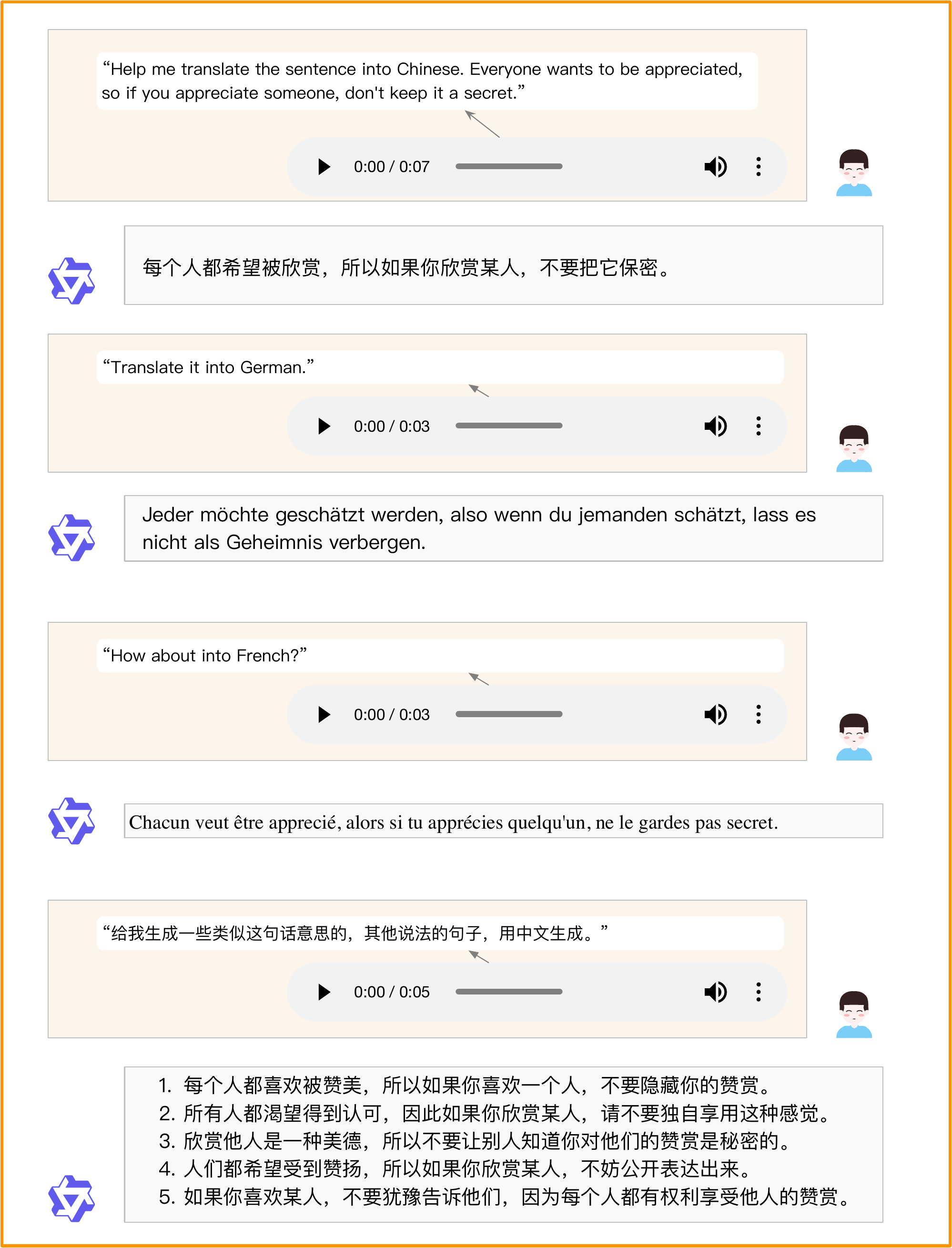}
\caption{Example showing Qwen2-Audio’s capability in free chat around speech.}
\end{figure*}

\begin{figure*}[htbp]
\centering
\includegraphics[width=0.7\textwidth]
    {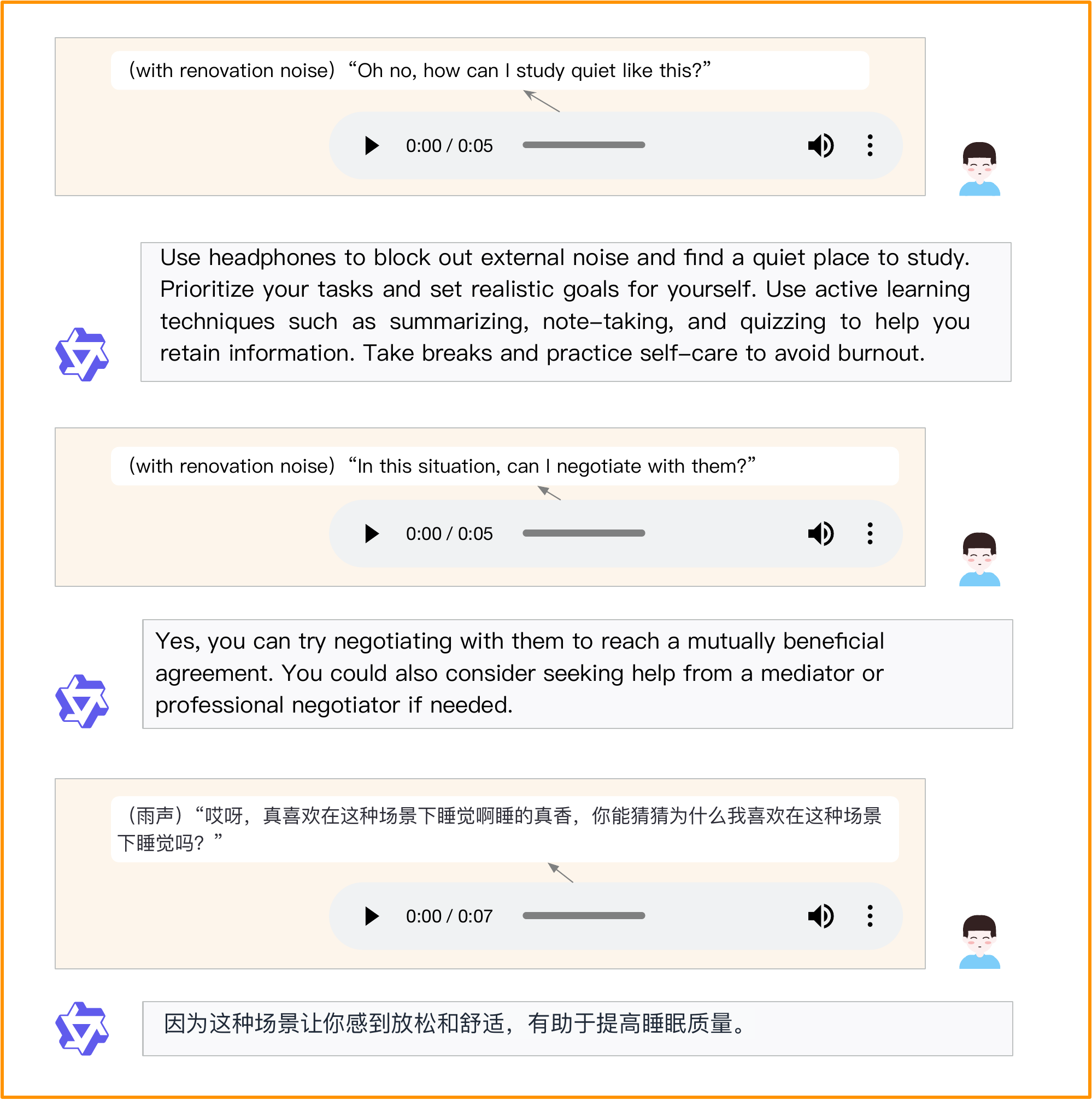}
\caption{Example showing Qwen2-Audio’s capability in free chat around speech and nature sound.}
\end{figure*}


\begin{figure*}[htbp]
\centering
\includegraphics[width=0.7\textwidth]
    {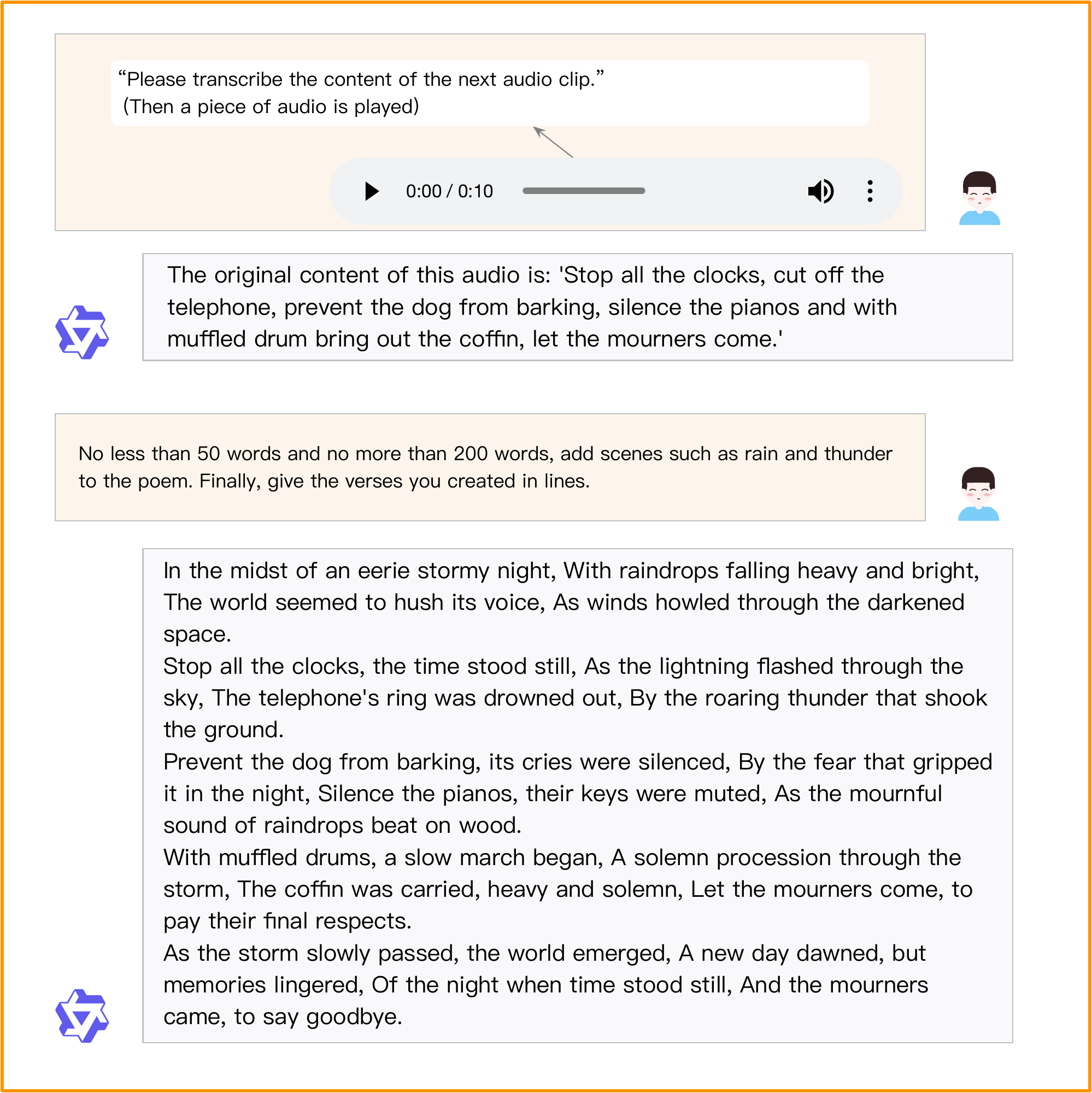}
\caption{Example showing Qwen2-Audio’s capability in speech analysis.}
\end{figure*}

\begin{figure*}[htbp]
\centering
\includegraphics[width=0.7\textwidth]
    {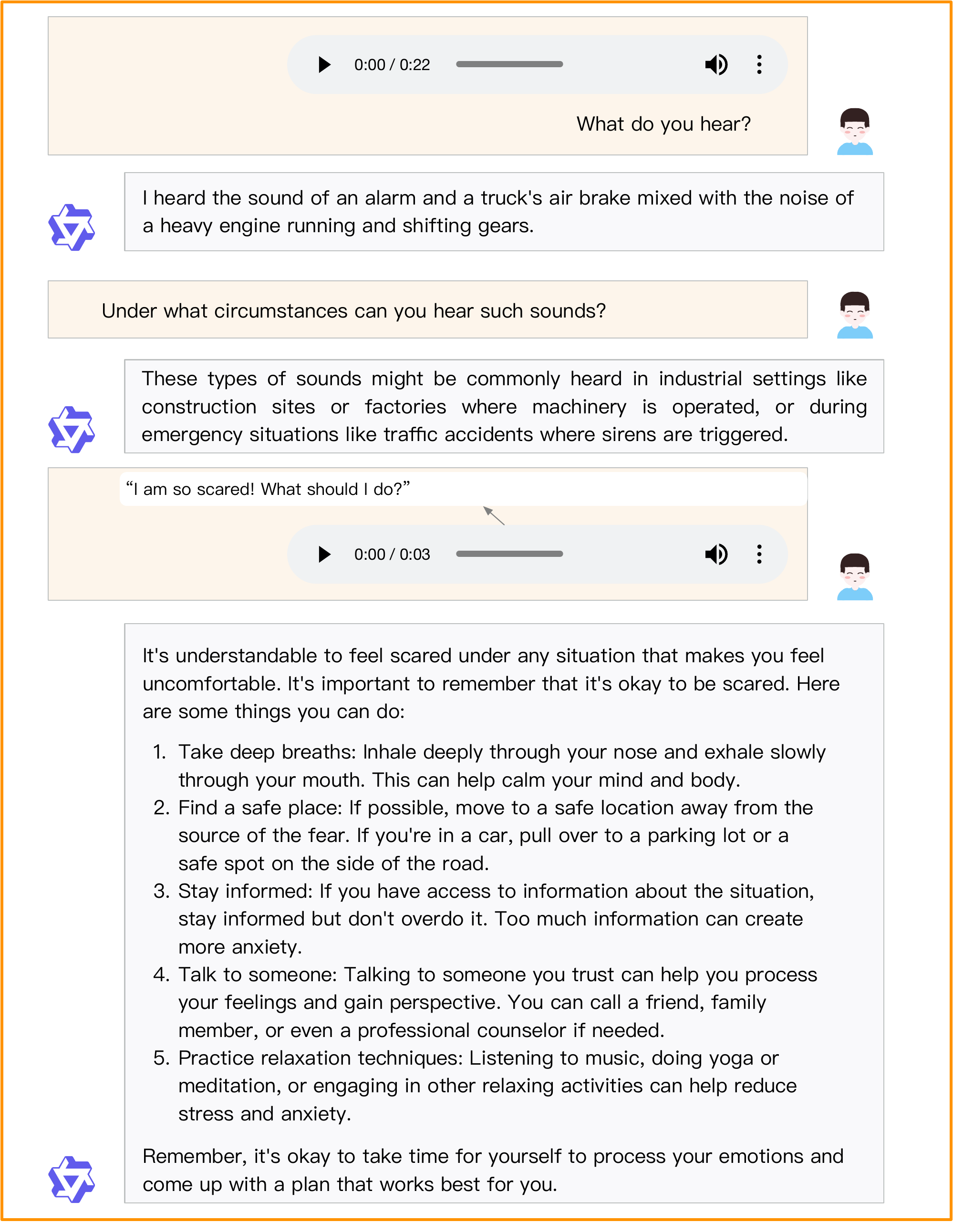}
\caption{Example showing Qwen2-Audio’s capability in sound analysis.}
\end{figure*}

\begin{figure*}[htbp]
\centering
\includegraphics[width=0.7\textwidth]
    {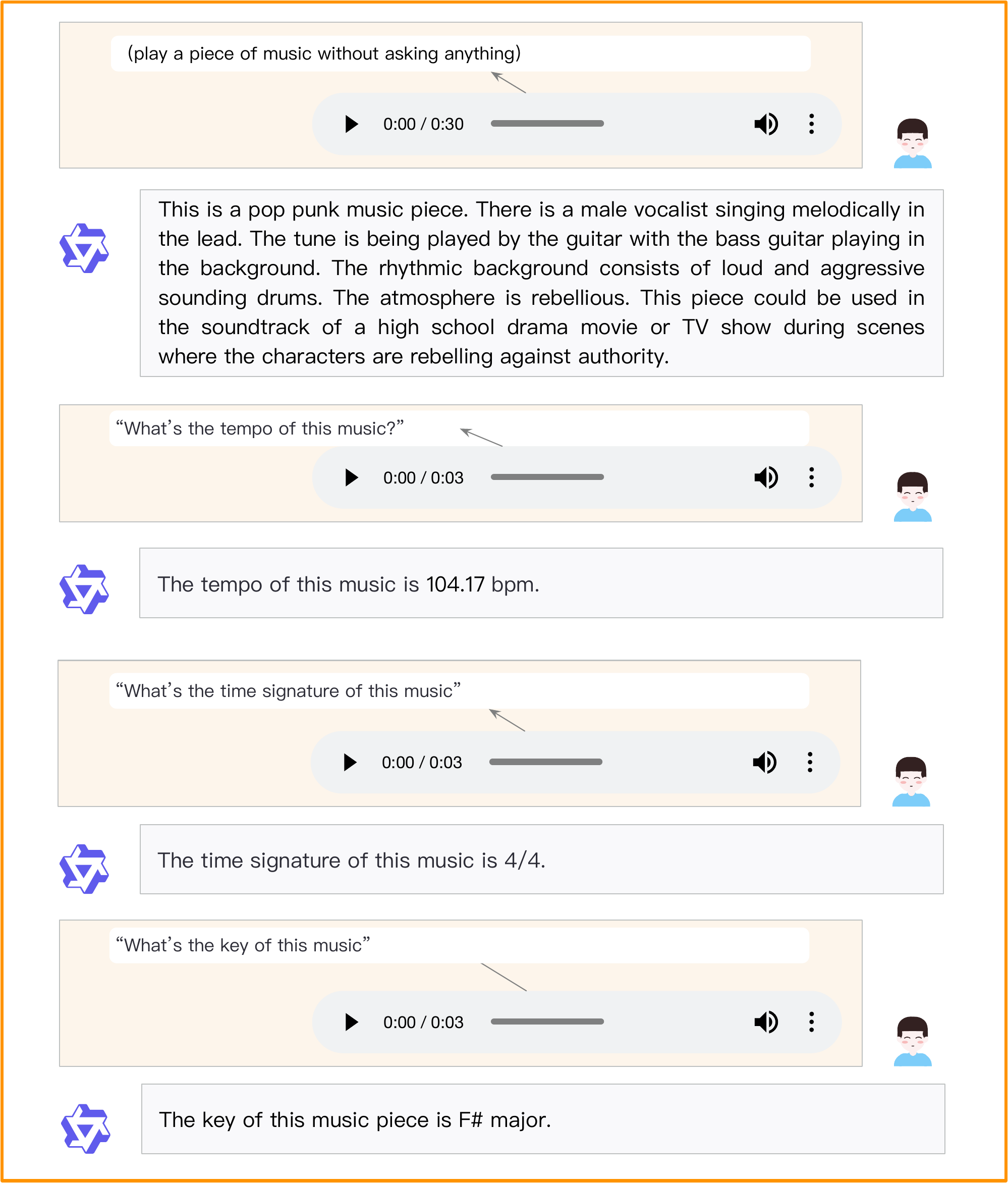}
\caption{Example showing Qwen2-Audio’s capability in music analysis.}
\end{figure*}

\begin{figure*}[htbp]
\centering
\includegraphics[width=0.7\textwidth]
    {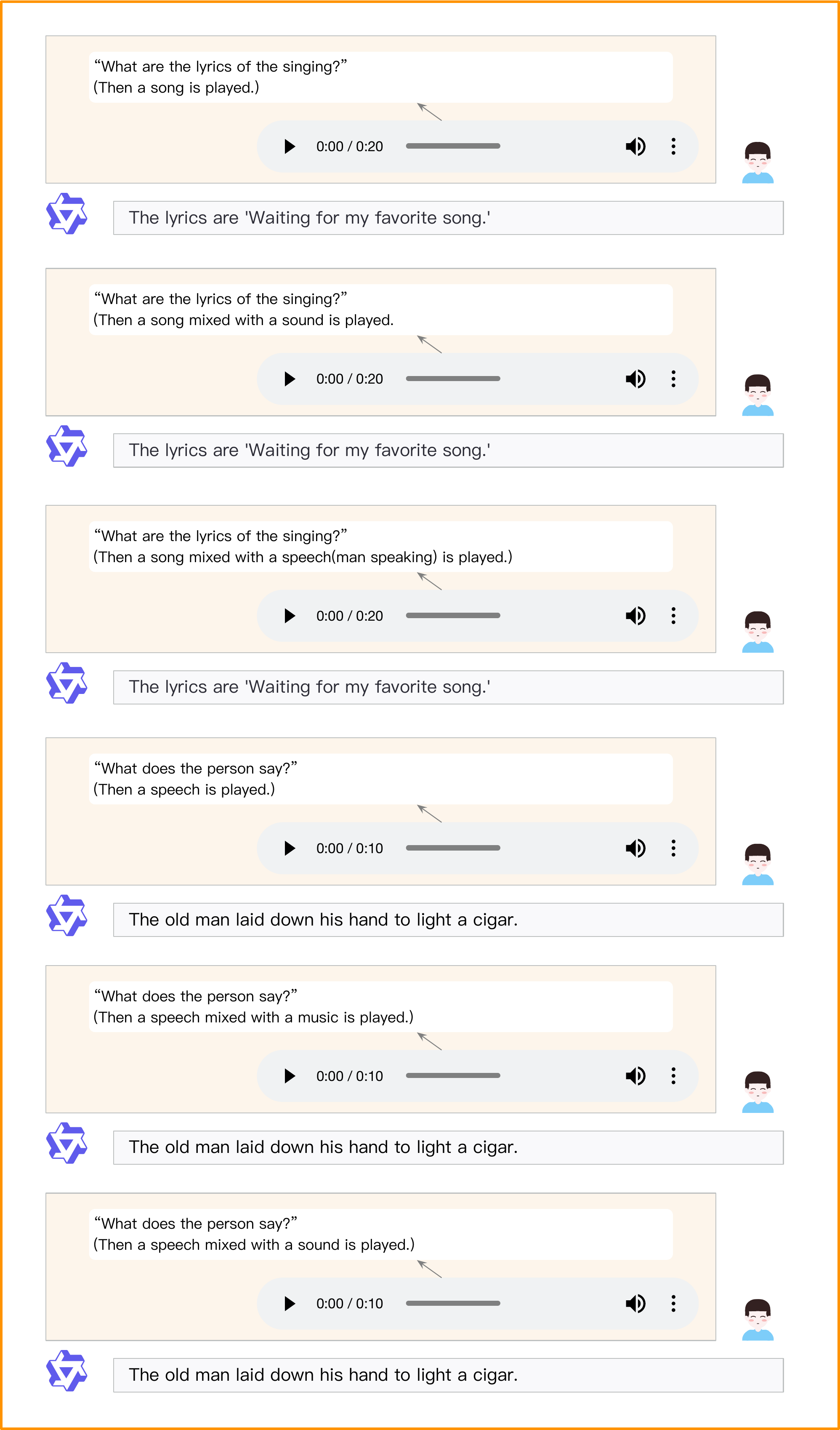}
\caption{Example showing Qwen2-Audio’s robustness in mixed audio analysis.}
\end{figure*}

\section{Conclusion}
\label{sec:conclusion}

In this paper, we present Qwen2-Audio, which builds upon Qwen-Audio's capability to analyze various types of audio while also being endowed with voice interaction abilities. During the pre-training stage, we utilized natural language prompts for different data and tasks and have further expanded the data volume. In the SFT phase, we enhanced Qwen2-Audio's alignment with human interaction by increasing the quantity, quality, and complexity of SFT data, thereby enabling seamless voice and text interactions. Additionally, we improved Qwen2-Audio's response quality through the DPO stage. Objective metrics tested on diverse benchmarks demonstrate Qwen2-Audio's proficiency in audio understanding and dialogue capabilities. The cases presented within the paper also illustrate Qwen2-Audio's fluent and flexible voice interaction capability.

\section{Acknowledgements}
We express our gratitude to Jinze Bai, Shuai Bai, Peng Wang, Sinan Tan, Shijie Wang, Kai Dang for their insightful discussion.

\bibliographystyle{plainnat} 
\bibliography{references}

\begin{thebibliography}{39}
\providecommand{\natexlab}[1]{#1}
\providecommand{\url}[1]{\texttt{#1}}
\expandafter\ifx\csname urlstyle\endcsname\relax
  \providecommand{\doi}[1]{doi: #1}\else
  \providecommand{\doi}{doi: \begingroup \urlstyle{rm}\Url}\fi

\bibitem[Agostinelli et~al.(2023)Agostinelli, Denk, Borsos, Engel, Verzetti, Caillon, Huang, Jansen, Roberts, Tagliasacchi, et~al.]{agostinelli2023musiclm}
Andrea Agostinelli, Timo~I Denk, Zal{\'a}n Borsos, Jesse Engel, Mauro Verzetti, Antoine Caillon, Qingqing Huang, Aren Jansen, Adam Roberts, Marco Tagliasacchi, et~al.
\newblock Musiclm: Generating music from text.
\newblock \emph{arXiv preprint arXiv:2301.11325}, 2023.

\bibitem[Ao et~al.(2021)Ao, Wang, Zhou, Wang, Ren, Wu, Liu, Ko, Li, Zhang, et~al.]{ao2021speecht5}
Junyi Ao, Rui Wang, Long Zhou, Chengyi Wang, Shuo Ren, Yu~Wu, Shujie Liu, Tom Ko, Qing Li, Yu~Zhang, et~al.
\newblock Speecht5: Unified-modal encoder-decoder pre-training for spoken language processing.
\newblock \emph{arXiv:2110.07205}, 2021.

\bibitem[Ardila et~al.(2020)Ardila, Branson, Davis, Henretty, Kohler, Meyer, Morais, Saunders, Tyers, and Weber]{commonvoice:2020}
R.~Ardila, M.~Branson, K.~Davis, M.~Henretty, M.~Kohler, J.~Meyer, R.~Morais, L.~Saunders, F.~M. Tyers, and G.~Weber.
\newblock Common voice: A massively-multilingual speech corpus.
\newblock In \emph{Proceedings of the 12th Conference on Language Resources and Evaluation (LREC 2020)}, pages 4211--4215, 2020.

\bibitem[Bai et~al.(2023)Bai, Bai, Chu, Cui, Dang, Deng, Fan, Ge, Han, Huang, et~al.]{bai2023qwen}
Jinze Bai, Shuai Bai, Yunfei Chu, Zeyu Cui, Kai Dang, Xiaodong Deng, Yang Fan, Wenbin Ge, Yu~Han, Fei Huang, et~al.
\newblock Qwen technical report.
\newblock \emph{arXiv preprint arXiv:2309.16609}, 2023.

\bibitem[Chen et~al.(2022)Chen, Wang, Chen, Wu, Liu, Chen, Li, Kanda, Yoshioka, Xiao, Wu, Zhou, Ren, Qian, Qian, Wu, Zeng, Yu, and Wei]{wavlm}
Sanyuan Chen, Chengyi Wang, Zhengyang Chen, Yu~Wu, Shujie Liu, Zhuo Chen, Jinyu Li, Naoyuki Kanda, Takuya Yoshioka, Xiong Xiao, Jian Wu, Long Zhou, Shuo Ren, Yanmin Qian, Yao Qian, Jian Wu, Michael Zeng, Xiangzhan Yu, and Furu Wei.
\newblock Wavlm: Large-scale self-supervised pre-training for full stack speech processing.
\newblock \emph{{IEEE} J. Sel. Top. Signal Process.}, 2022.

\bibitem[Chen et~al.(2021)Chen, Chi, Yang, Chang, Lin, Huang, Liu, Liu, Lee, and Lee]{chen2021speechnet}
Yi-Chen Chen, Po-Han Chi, Shu-wen Yang, Kai-Wei Chang, Jheng-hao Lin, Sung-Feng Huang, Da-Rong Liu, Chi-Liang Liu, Cheng-Kuang Lee, and Hung-yi Lee.
\newblock Speechnet: A universal modularized model for speech processing tasks.
\newblock \emph{arXiv:2105.03070}, 2021.

\bibitem[Chu et~al.(2023)Chu, Xu, Zhou, Yang, Zhang, Yan, Zhou, and Zhou]{chu2023qwen}
Yunfei Chu, Jin Xu, Xiaohuan Zhou, Qian Yang, Shiliang Zhang, Zhijie Yan, Chang Zhou, and Jingren Zhou.
\newblock Qwen-audio: Advancing universal audio understanding via unified large-scale audio-language models.
\newblock \emph{arXiv preprint arXiv:2311.07919}, 2023.

\bibitem[Cieri et~al.(2004)Cieri, Miller, and Walker]{cieri2004fisher}
Christopher Cieri, David Miller, and Kevin Walker.
\newblock The fisher corpus: A resource for the next generations of speech-to-text.
\newblock In \emph{LREC}, volume~4, pages 69--71, 2004.

\bibitem[Conneau et~al.(2022)Conneau, Ma, Khanuja, Zhang, Axelrod, Dalmia, Riesa, Rivera, and Bapna]{Conneau2022FLEURSFL}
Alexis Conneau, Min Ma, Simran Khanuja, Yu~Zhang, Vera Axelrod, Siddharth Dalmia, Jason Riesa, Clara Rivera, and Ankur Bapna.
\newblock Fleurs: Few-shot learning evaluation of universal representations of speech.
\newblock \emph{2022 IEEE Spoken Language Technology Workshop (SLT)}, pages 798--805, 2022.
\newblock URL \url{https://api.semanticscholar.org/CorpusID:249062909}.

\bibitem[Das et~al.(2024)Das, Dingliwal, Ronanki, Paturi, Huang, Mathur, Yuan, Bekal, Niu, Jayanthi, et~al.]{das2024speechverse}
Nilaksh Das, Saket Dingliwal, Srikanth Ronanki, Rohit Paturi, David Huang, Prashant Mathur, Jie Yuan, Dhanush Bekal, Xing Niu, Sai~Muralidhar Jayanthi, et~al.
\newblock Speechverse: A large-scale generalizable audio language model.
\newblock \emph{arXiv preprint arXiv:2405.08295}, 2024.

\bibitem[Deshmukh et~al.(2023)Deshmukh, Elizalde, Singh, and Wang]{Pengi}
Soham Deshmukh, Benjamin Elizalde, Rita Singh, and Huaming Wang.
\newblock Pengi: An audio language model for audio tasks.
\newblock \emph{CoRR}, 2023.

\bibitem[Drossos et~al.(2020)Drossos, Lipping, and Virtanen]{Clotho}
Konstantinos Drossos, Samuel Lipping, and Tuomas Virtanen.
\newblock Clotho: an audio captioning dataset.
\newblock In \emph{2020 {IEEE} International Conference on Acoustics, Speech and Signal Processing, {ICASSP} 2020, Barcelona, Spain, May 4-8, 2020}. {IEEE}, 2020.

\bibitem[Du et~al.(2018)Du, Na, Liu, and Bu]{aishell2}
Jiayu Du, Xingyu Na, Xuechen Liu, and Hui Bu.
\newblock {AISHELL-2:} transforming mandarin {ASR} research into industrial scale.
\newblock abs/1808.10583, 2018.

\bibitem[Elizalde et~al.(2022)Elizalde, Deshmukh, Ismail, and Wang]{CLAP}
Benjamin Elizalde, Soham Deshmukh, Mahmoud~Al Ismail, and Huaming Wang.
\newblock {CLAP:} learning audio concepts from natural language supervision.
\newblock abs/2206.04769, 2022.

\bibitem[Gao et~al.(2023)Gao, Li, Wang, Luo, Shi, Chen, Li, Zuo, Du, Xiao, and Zhang]{FunASR}
Zhifu Gao, Zerui Li, Jiaming Wang, Haoneng Luo, Xian Shi, Mengzhe Chen, Yabin Li, Lingyun Zuo, Zhihao Du, Zhangyu Xiao, and Shiliang Zhang.
\newblock Funasr: {A} fundamental end-to-end speech recognition toolkit.
\newblock \emph{CoRR}, abs/2305.11013, 2023.

\bibitem[Gong et~al.(2022)Gong, Yu, and Glass]{VocalSound}
Yuan Gong, Jin Yu, and James~R. Glass.
\newblock Vocalsound: {A} dataset for improving human vocal sounds recognition.
\newblock In \emph{{IEEE} International Conference on Acoustics, Speech and Signal Processing, {ICASSP} 2022, Virtual and Singapore, 23-27 May 2022}, pages 151--155. {IEEE}, 2022.
\newblock \doi{10.1109/ICASSP43922.2022.9746828}.
\newblock URL \url{https://doi.org/10.1109/ICASSP43922.2022.9746828}.

\bibitem[Kim et~al.(2019)Kim, Kim, Lee, and Kim]{kim2019audiocaps}
Chris~Dongjoo Kim, Byeongchang Kim, Hyunmin Lee, and Gunhee Kim.
\newblock Audiocaps: Generating captions for audios in the wild.
\newblock In \emph{Proceedings of the 2019 Conference of the North American Chapter of the Association for Computational Linguistics: Human Language Technologies, Volume 1 (Long and Short Papers)}, 2019.

\bibitem[Kong et~al.(2024)Kong, Goel, Badlani, Ping, Valle, and Catanzaro]{kong2024audio}
Zhifeng Kong, Arushi Goel, Rohan Badlani, Wei Ping, Rafael Valle, and Bryan Catanzaro.
\newblock Audio flamingo: A novel audio language model with few-shot learning and dialogue abilities.
\newblock \emph{arXiv preprint arXiv:2402.01831}, 2024.

\bibitem[Lyu et~al.(2023)Lyu, Wu, Wang, Huang, Liu, Du, Shi, and Tu]{Macaw-LLM}
Chenyang Lyu, Minghao Wu, Longyue Wang, Xinting Huang, Bingshuai Liu, Zefeng Du, Shuming Shi, and Zhaopeng Tu.
\newblock Macaw-llm: Multi-modal language modeling with image, audio, video, and text integration.
\newblock \emph{CoRR}, abs/2306.09093, 2023.

\bibitem[OpenAI(2023)]{gpt4}
OpenAI.
\newblock Gpt-4 technical report, 2023.

\bibitem[OpenAI(2024)]{openai2024gpt4-o}
OpenAI.
\newblock Gpt-4o, 2024.
\newblock URL \url{https://openai.com/index/hello-gpt-4o/}.

\bibitem[Panayotov et~al.(2015)Panayotov, Chen, Povey, and Khudanpur]{Librispeech}
Vassil Panayotov, Guoguo Chen, Daniel Povey, and Sanjeev Khudanpur.
\newblock Librispeech: An {ASR} corpus based on public domain audio books.
\newblock In \emph{2015 {IEEE} International Conference on Acoustics, Speech and Signal Processing, {ICASSP} 2015, South Brisbane, Queensland, Australia, April 19-24, 2015}. {IEEE}, 2015.

\bibitem[Papineni et~al.(2002)Papineni, Roukos, Ward, and Zhu]{papineni2002bleu}
Kishore Papineni, Salim Roukos, Todd Ward, and Wei-Jing Zhu.
\newblock Bleu: a method for automatic evaluation of machine translation.
\newblock In \emph{Proceedings of the 40th annual meeting of the Association for Computational Linguistics}, 2002.

\bibitem[Poria et~al.(2019)Poria, Hazarika, Majumder, Naik, Cambria, and Mihalcea]{Meld}
Soujanya Poria, Devamanyu Hazarika, Navonil Majumder, Gautam Naik, Erik Cambria, and Rada Mihalcea.
\newblock {MELD:} {A} multimodal multi-party dataset for emotion recognition in conversations.
\newblock In \emph{Proceedings of the 57th Conference of the Association for Computational Linguistics, {ACL} 2019, Florence, Italy, July 28- August 2, 2019, Volume 1: Long Papers}. Association for Computational Linguistics, 2019.

\bibitem[Qwen(2023)]{qwen7b}
Qwen.
\newblock Introducing qwen-7b: Open foundation and human-aligned models (of the state-of-the-arts), 2023.
\newblock URL \url{https://github.com/QwenLM/Qwen-7B}.

\bibitem[Radford et~al.(2023)Radford, Kim, Xu, Brockman, McLeavey, and Sutskever]{Whisper}
Alec Radford, Jong~Wook Kim, Tao Xu, Greg Brockman, Christine McLeavey, and Ilya Sutskever.
\newblock Robust speech recognition via large-scale weak supervision.
\newblock In \emph{International Conference on Machine Learning, {ICML} 2023, 23-29 July 2023, Honolulu, Hawaii, {USA}}, 2023.

\bibitem[Rafailov et~al.(2024)Rafailov, Sharma, Mitchell, Manning, Ermon, and Finn]{rafailov2024direct}
Rafael Rafailov, Archit Sharma, Eric Mitchell, Christopher~D Manning, Stefano Ermon, and Chelsea Finn.
\newblock Direct preference optimization: Your language model is secretly a reward model.
\newblock \emph{Advances in Neural Information Processing Systems}, 36, 2024.

\bibitem[Reid et~al.(2024)Reid, Savinov, Teplyashin, Lepikhin, Lillicrap, Alayrac, Soricut, Lazaridou, Firat, Schrittwieser, et~al.]{reid2024gemini}
Machel Reid, Nikolay Savinov, Denis Teplyashin, Dmitry Lepikhin, Timothy Lillicrap, Jean-baptiste Alayrac, Radu Soricut, Angeliki Lazaridou, Orhan Firat, Julian Schrittwieser, et~al.
\newblock Gemini 1.5: Unlocking multimodal understanding across millions of tokens of context.
\newblock \emph{arXiv preprint arXiv:2403.05530}, 2024.

\bibitem[Si et~al.(2023)Si, Ma, Wu, Dai, Gao, Lin, Li, Yan, Huang, and Li]{si2023spokenwoz}
Shuzheng Si, Wentao Ma, Yuchuan Wu, Yinpei Dai, Haoyu Gao, Ting-En Lin, Hangyu Li, Rui Yan, Fei Huang, and Yongbin Li.
\newblock Spokenwoz: A large-scale speech-text benchmark for spoken task-oriented dialogue in multiple domains.
\newblock \emph{arXiv preprint arXiv:2305.13040}, 2023.

\bibitem[Su et~al.(2023)Su, Lan, Li, Xu, Wang, and Cai]{su2023pandagpt}
Yixuan Su, Tian Lan, Huayang Li, Jialu Xu, Yan Wang, and Deng Cai.
\newblock Pandagpt: One model to instruction-follow them all.
\newblock \emph{arXiv:2305.16355}, 2023.

\bibitem[Tang et~al.(2024)Tang, Yu, Sun, Chen, Tan, Li, Lu, MA, and Zhang]{anonymous2023salmonn}
Changli Tang, Wenyi Yu, Guangzhi Sun, Xianzhao Chen, Tian Tan, Wei Li, Lu~Lu, Zejun MA, and Chao Zhang.
\newblock {SALMONN}: Towards generic hearing abilities for large language models.
\newblock In \emph{The Twelfth International Conference on Learning Representations}, 2024.
\newblock URL \url{https://openreview.net/forum?id=14rn7HpKVk}.

\bibitem[Wang et~al.(2020)Wang, Wu, and Pino]{CoVoST2}
Changhan Wang, Anne Wu, and Juan~Miguel Pino.
\newblock Covost 2: {A} massively multilingual speech-to-text translation corpus.
\newblock abs/2007.10310, 2020.
\newblock URL \url{https://arxiv.org/abs/2007.10310}.

\bibitem[Wang et~al.(2023{\natexlab{a}})Wang, Liao, Huang, Lu, Wu, Liu, Zong, and Zhang]{wang2023blsp}
Chen Wang, Minpeng Liao, Zhongqiang Huang, Jinliang Lu, Junhong Wu, Yuchen Liu, Chengqing Zong, and Jiajun Zhang.
\newblock Blsp: Bootstrapping language-speech pre-training via behavior alignment of continuation writing.
\newblock \emph{arXiv:2309.00916}, 2023{\natexlab{a}}.

\bibitem[Wang et~al.(2023{\natexlab{b}})Wang, Han, Shafran, Wu, Chiu, Cao, Wang, Chen, Zhang, Soltau, Rubenstein, Zilka, Yu, Meng, Pundak, Siddhartha, Schalkwyk, and Wu]{SLM}
Mingqiu Wang, Wei Han, Izhak Shafran, Zelin Wu, Chung{-}Cheng Chiu, Yuan Cao, Yongqiang Wang, Nanxin Chen, Yu~Zhang, Hagen Soltau, Paul~K. Rubenstein, Lukas Zilka, Dian Yu, Zhong Meng, Golan Pundak, Nikhil Siddhartha, Johan Schalkwyk, and Yonghui Wu.
\newblock {SLM:} bridge the thin gap between speech and text foundation models.
\newblock abs/2310.00230, 2023{\natexlab{b}}.

\bibitem[Wu et~al.(2023{\natexlab{a}})Wu, Gaur, Chen, Zhou, Zhu, Wang, Li, Liu, Ren, Liu, and Wu]{speechllama}
Jian Wu, Yashesh Gaur, Zhuo Chen, Long Zhou, Yimeng Zhu, Tianrui Wang, Jinyu Li, Shujie Liu, Bo~Ren, Linquan Liu, and Yu~Wu.
\newblock On decoder-only architecture for speech-to-text and large language model integration.
\newblock abs/2307.03917, 2023{\natexlab{a}}.

\bibitem[Wu et~al.(2023{\natexlab{b}})Wu, Fei, Qu, Ji, and Chua]{NextGPT}
Shengqiong Wu, Hao Fei, Leigang Qu, Wei Ji, and Tat{-}Seng Chua.
\newblock Next-gpt: Any-to-any multimodal {LLM}.
\newblock \emph{CoRR}, abs/2309.05519, 2023{\natexlab{b}}.

\bibitem[Yang et~al.(2024)Yang, Xu, Liu, Chu, Jiang, Zhou, Leng, Lv, Zhao, Zhou, and Zhou]{Yang2024AIRBenchBL}
Qian Yang, Jin Xu, Wenrui Liu, Yunfei Chu, Ziyue Jiang, Xiaohuan Zhou, Yichong Leng, Yuanjun Lv, Zhou Zhao, Chang Zhou, and Jingren Zhou.
\newblock Air-bench: Benchmarking large audio-language models via generative comprehension.
\newblock In \emph{ACL}, 2024.

\bibitem[Zhang et~al.(2023)Zhang, Li, Zhang, Zhan, Wang, Zhou, and Qiu]{speechgpt}
Dong Zhang, Shimin Li, Xin Zhang, Jun Zhan, Pengyu Wang, Yaqian Zhou, and Xipeng Qiu.
\newblock Speechgpt: Empowering large language models with intrinsic cross-modal conversational abilities.
\newblock \emph{CoRR}, abs/2305.11000, 2023.

\bibitem[Zhou et~al.(2022)Zhou, Wang, Cui, Zhang, Yan, Zhou, and Zhou]{mmspeech}
Xiaohuan Zhou, Jiaming Wang, Zeyu Cui, Shiliang Zhang, Zhijie Yan, Jingren Zhou, and Chang Zhou.
\newblock Mmspeech: Multi-modal multi-task encoder-decoder pre-training for speech recognition.
\newblock abs/2212.00500, 2022.

\end{thebibliography}




\end{document}